\documentclass[conference]{IEEEtran}
\IEEEoverridecommandlockouts
\usepackage{cite}
\usepackage{amsmath,amssymb,amsfonts}
\usepackage{algorithmic}
\usepackage{graphicx}
\usepackage{textcomp}
\usepackage{xcolor}
\usepackage{subcaption}
\usepackage{hyperref}
\usepackage[utf8]{inputenc}
\usepackage{multirow}

\def\BibTeX{{\rm B\kern-.05em{\sc i\kern-.025em b}\kern-.08em
    T\kern-.1667em\lower.7ex\hbox{E}\kern-.125emX}}
\begin{document}

\title{GPS-IMU Sensor Fusion for Reliable Autonomous Vehicle Position Estimation\\
}

\author{\IEEEauthorblockN{Simegnew Yihunie Alaba 
}
\IEEEauthorblockA{\textit{Department of Electrical and Computer Engineering} \\
\textit{Mississippi State University}\\
Starkville, United States \\
}
}
\maketitle

\begin{abstract}
Global Positioning System (GPS) navigation provides accurate positioning with global coverage, making it a reliable option in open areas with unobstructed sky views. However, signal degradation may occur in indoor spaces and urban canyons. In contrast, Inertial Measurement Units (IMUs) consist of gyroscopes and accelerometers that offer relative motion information such as acceleration and rotational changes. Unlike GPS, IMUs do not rely on external signals, making them useful in GPS-denied environments. Nonetheless, IMUs suffer from drift over time due to the accumulation of errors while integrating acceleration to determine velocity and position. Therefore, fusing the GPS and IMU is crucial for enhancing the reliability and precision of navigation systems in autonomous vehicles, especially in environments where GPS signals are compromised. To ensure smooth navigation and overcome the limitations of each sensor, the proposed method fuses GPS and IMU data. This sensor fusion uses the Unscented Kalman Filter (UKF) Bayesian filtering technique. The proposed navigation system is designed to be robust, delivering continuous and accurate positioning critical for the safe operation of autonomous vehicles, particularly in GPS-denied environments. This project uses KITTI GNSS and IMU datasets for experimental validation, showing that the GNSS-IMU fusion technique reduces GNSS-only data's RMSE. The RMSE decreased from 13.214, 13.284, and 13.363 to 4.271, 5.275, and 0.224 for the x-axis, y-axis, and z-axis, respectively. The experimental result using UKF shows promising direction in improving autonomous vehicle navigation using GPS and IMU sensor fusion using the best of two sensors in GPS-denied environments.

\end{abstract}


\section{Introduction}
In autonomous vehicle navigation, integrating Global Positioning System (GPS) and Inertial Measurement Units (IMU) has become a cornerstone for achieving reliable and precise location tracking, particularly in challenging environments. While GPS offers extensive coverage and high positional accuracy in open spaces  \cite{bevly2004global}, its performance degrades in indoor or urban canyons where signals are obstructed \cite{li2014reliable}. Conversely, IMUs provide valuable motion data independently of external signals, making them indispensable in GPS-denied areas. However, the utility of IMUs is hindered by their susceptibility to drift over time, which accumulates errors in velocity and position estimations derived from acceleration data.

To mitigate the limitations of each sensor type, the fusion of GPS and IMU data emerges as a crucial strategy. This fusion aims to leverage the global positioning capabilities of GPS with the relative motion insights from IMUs, thus enhancing the robustness and accuracy of navigation systems in autonomous vehicles. The application of advanced Bayesian filtering techniques, particularly the Extended Kalman Filter (EKF) \cite{sorenson1966kalman}  and Unscented Kalman Filter (UKF) \cite{wan2000unscented}, facilitates the effective integration of these sensors. This approach ensures seamless and reliable navigation, which is vital for the reliable operation of autonomous vehicles, especially in environments where GPS signals are compromised.

This project, which utilizes the KITTI GNSS and IMU datasets for validation, demonstrates its potential through realistic experimental setups. These initiatives contribute to the development of the technological infrastructure for self-driving vehicles and tackle the crucial safety and efficacy issues facing the industry at present.

\section{Literature Review}

The GPS and IMU fusion is essential for autonomous vehicle navigation. It addresses limitations when these sensors operate independently, particularly in environments with weak or obstructed GPS signals, such as urban areas or indoor settings. Given the rising demand for robust autonomous navigation, developing sensor fusion methodologies that ensure reliable vehicle navigation is essential.

Various filtering techniques are used to integrate GNSS/GPS and IMU data effectively, with Kalman Filters \cite{kalman1960contributions} and their variants, such as the Extended Kalman Filter (EKF), the Unscented Kalman Filter (UKF), etc. Caron {\em et al.} \cite{caron2006gps} introduced a multisensor Kalman filter technique incorporating contextual variables to improve GPS/IMU fusion reliability, especially in signal-distorted environments. Lee {\em et al.} \cite{lee2016camera} put forth a sensor fusion method that combines camera, GPS, and IMU data, utilizing an EKF to improve state estimation in GPS-denied scenarios. Different innovative sensor fusion methods push the boundaries of autonomous vehicle navigation. Suwandi {\em et al.} \cite{suwandi2017low} demonstrated a cost-effective approach to vehicle navigation by focusing on low-cost IMU and GPS sensor fusion to improve navigation. Atia {\em et al.} \cite{atia2017low} combined MEMS, IMU, GPS, and road network maps with an EKF and Hidden Markov model-based map-matching to provide accurate lane determination without high-precision GNSS technologies. Li and Xu \cite{li2016reliable} introduced a method for sensor fusion navigation in GPS-denied areas. This approach fuses cheaper sensors with a sliding mode observer and a federated Kalman filter.

Liu {\em et al.} \cite{liu2018innovative} developed an enhanced Adaptive Kalman Filter (IAE-AKF) with an attenuation factor to handle noise effectively, resulting in a 20\% improvement in navigation accuracy. Meng {\em et al.} \cite{meng2017robust} employed the Global Navigation Satellite System (GNSS), IMU, DMI, and LiDAR to counteract inaccuracies caused by GNSS signal jumps and multi-path interference in urban settings. Tao {\em et al.} \cite{tao2021multi} introduced a multisensor fusion strategy that integrates GNSS, IMU, and visual data with global pose graph optimization, yielding superior results on a ROS simulation platform using the KITTI dataset. Yusef {\em et al.} \cite{yusefi2023generalizable} employed a Deep Visual Inertial Odometry framework to improve accuracy in low-GNSS areas. At the same time, Park \cite{park2024optimal} used an adaptive Kalman filter for vehicle position estimation to address GPS outages, validated in real-world tests. Gruyer and Pollard \cite{gruyer2011credibilistic} enhanced navigation in GPS-denied environments using proprioceptive sensors with an Interacting Multiple Model (IMM) filter. Godoy {\em et al.} \cite{godoy2012development} proposed a Particle Swarm Optimization-based filter integrating multiple sensors with digital maps, showing promise in reducing reliance on GPS in accurate vehicle testing.

Although these studies propose novel methods to improve autonomous vehicle navigation, most rely on locally collected data, making their results difficult to reproduce. This work addresses this issue using a publicly available dataset, allowing others to improve autonomous driving navigation with real, representative, and accessible data. This proposed fusion technique leverages the strengths of both GNSS and IMU to maintain continuous operation, even if one sensor fails. It uses the publicly accessible KITTI dataset for testing, allowing others to replicate and validate the results. By experimenting with GPS-only and fused data, this study demonstrates the importance of sensor fusion for safe navigation in real-world conditions. This approach bridges the gap between theoretical models and practical applications, experimenting with a dataset representative of the real driving scenario.

\section{Method}
\begin{figure}[!ht]
    \centering
    \includegraphics[width=3.5in]{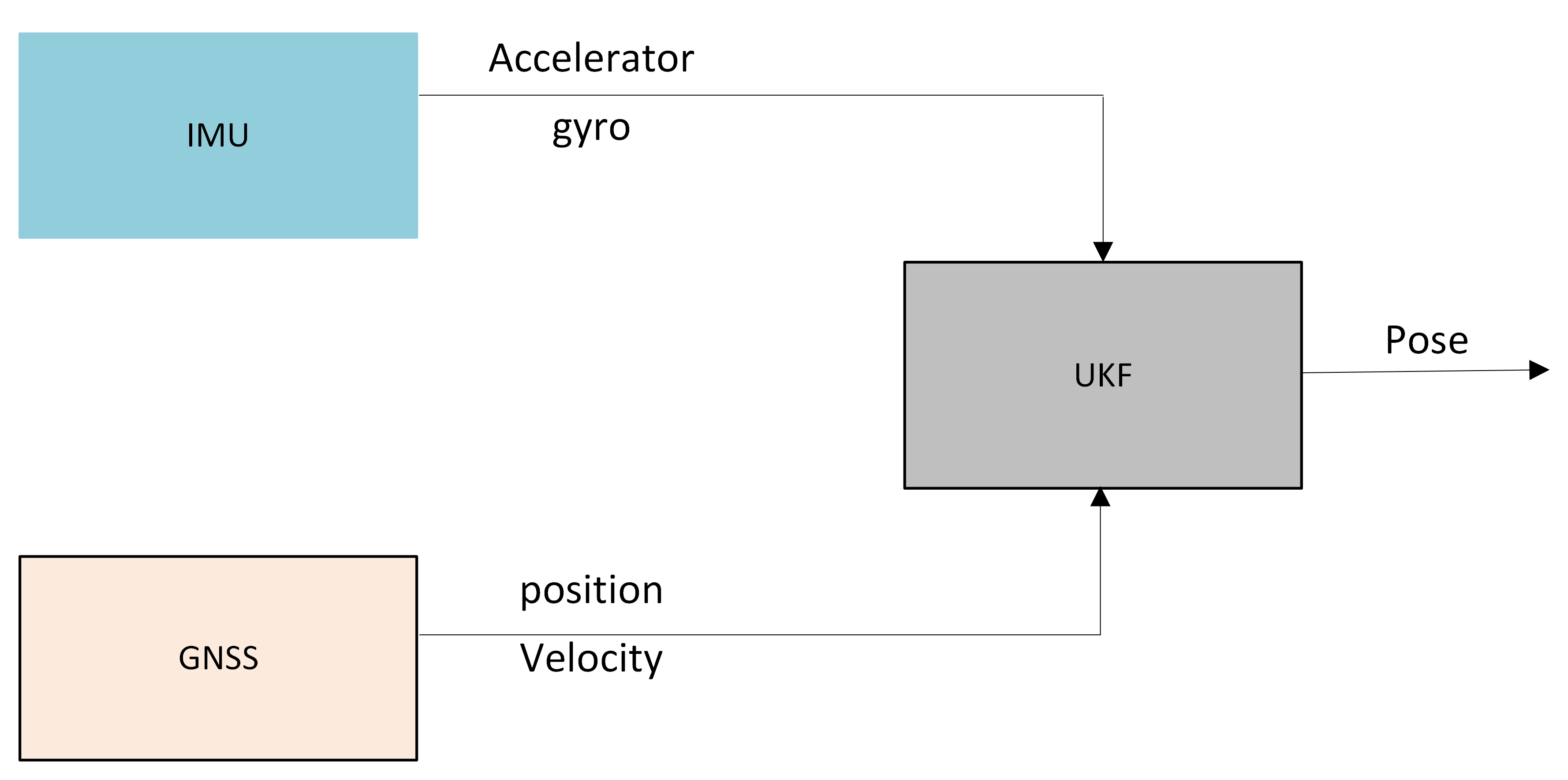}
    \caption{The proposed architecture.}
    \label{fig:proposed}
\end{figure}

Fig. \ref{fig:proposed} shows the proposed sensor fusion model for autonomous vehicle navigation. It fuses data from two primary sources: an IMU and a GNSS. With its accelerometers and gyroscopes, the IMU provides real-time information about the vehicle's acceleration and rotational movements, offering continuous data crucial for navigating without relying on external signals. 

On the other hand, GNSS offers absolute positioning and velocity data that are essential for high-precision location tracking when the vehicle has a clear path to the sky.
The UKF fuses the data from the two sensors. Unlike simpler models, the UKF is uniquely suited for handling the non-linear nature of the data integration challenge posed by autonomous vehicle navigation. By processing the high-rate, raw motion data from the IMU and the periodic positional corrections from the GNSS, the UKF corrects for any potential drift from the IMU and enhances its output, providing an accurate estimate of the vehicle's 'Pose'—its precise location and orientation in space.
This pose estimate is not merely a set of coordinates; it's a dynamic vehicle's position representation. It is continuously updated and adjusted, ensuring the vehicle 'knows' where it is, where it needs to go, and how to get there safely. This is vital for the vehicle's decision-making processes, such as navigating complex routes, avoiding obstacles, and ensuring passenger safety. The proposed work illustrates an equilibrium between robustness and accuracy, aiming to achieve seamless navigation functionality, particularly in demanding settings where GNSS signals may not be dependable or accessible.

The mathematical representation of the system comprises process and measurement models. The dynamic connection between the states at two consecutive time steps is managed by the process truth model, which can be defined as follows:

\begin{equation}
x_t = f\left( x_{t-1}, u_{t-1} \right) + w_{t-1},
\end{equation}
where \(x_{t}\) indicates the estimated state after an interval \(\sigma\), derived from the preceding state vector \(x_{t-1}\). The variable \(u_{t-1}\) functions as the input to the state space equations, and \(w_{t-1}\) reflects the noise affecting the process.

Data from the GNSS receiver introduce errors in the vehicle's position, velocity, and orientation estimates. When a measurement from the GNSS receiver is obtained, the GNSS measurement model is defined as:
\begin{equation}
    y_p = p + n_p,
\end{equation}
 where \( p \in (x, y, z) \), \( n_{p} \) represents the  measurement noise and $y_p = C_E^G[x^e, y^e, z^e]$.  Here, ($x^e, y^e, z^e$) are Earth-centered Earth-fixed (ECEF) rectangular coordinates \cite{meng2017robust,noureldin2012fundamentals}.
\begin{equation}
\begin{bmatrix}
x^e \\
y^e \\
z^e
\end{bmatrix}
=
\begin{bmatrix}
(R_N + h) \cos \phi \cos \lambda \\
(R_N + h) \cos \phi \sin \lambda \\
\left[ R_N(1 - e^2) + h \right] \sin \phi
\end{bmatrix}
\end{equation}
\text{where} \\
\begin{align*}
R_N & \text{ is the normal radius} \\
h & \text{ is the ellipsoidal height} \\
\lambda & \text{ is the longitude} \\
\phi & \text{ is the latitude} \\
e & \text{ is the eccentricity}.
\end{align*}

The normal radius $R_{N}$ is defined as: $R_N = \frac{a}{\sqrt{1 - e^2 \sin^2 \phi}}$. The Earth's eccentricity can also be expressed as \( e = \sqrt{\frac{a^2 - b^2}{a^2}} \) where \(a = 6,378,137 \, \text{m}\) and \(b = 6,356,752.3142 \, \text{m}\) are the semi-major and semi-minor axes of the Earth's ellipsoid, respectively.
The transition matrix from the ECEF frame to the global frame  {\em G}, denoted by  $C_{E}^{G}$,  is given as \cite{noureldin2012fundamentals}:

\begin{equation}
C_E^G =
\begin{bmatrix}
    -\sin \lambda \cos \phi & -\sin \lambda \sin \phi & \cos \lambda \\
    -\sin \phi & -\cos \phi & 0 \\
    -\cos \lambda \cos \phi & -\cos \lambda \sin \phi & -\sin \lambda
\end{bmatrix}.
\end{equation}
The covariance matrix of the GNSS measurement noise is $ \mathbf{R} = \Sigma_{\text{GNSS}}$.

\begin{figure*}[!ht]
    \centering
    \begin{tabular}{ccc}
     \includegraphics[width=0.3\textwidth]{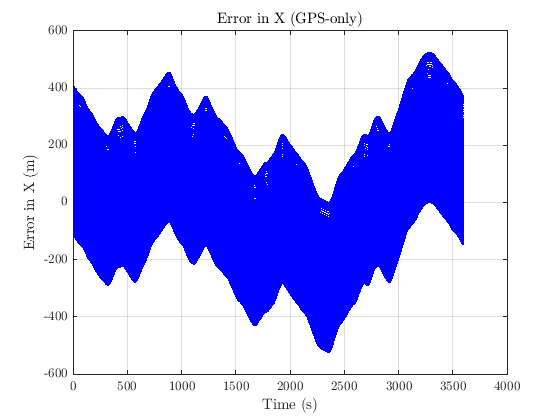} &
    \includegraphics[width=0.3\textwidth]{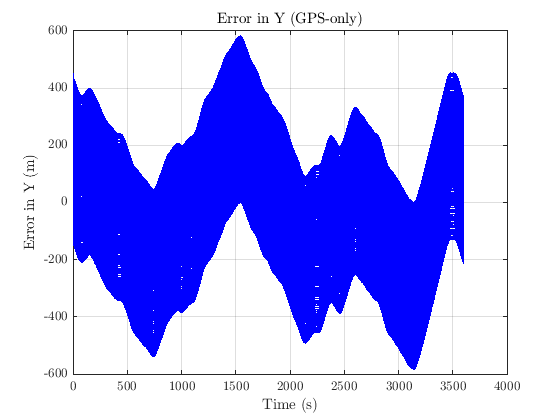} &
    \includegraphics[width=0.3\textwidth]{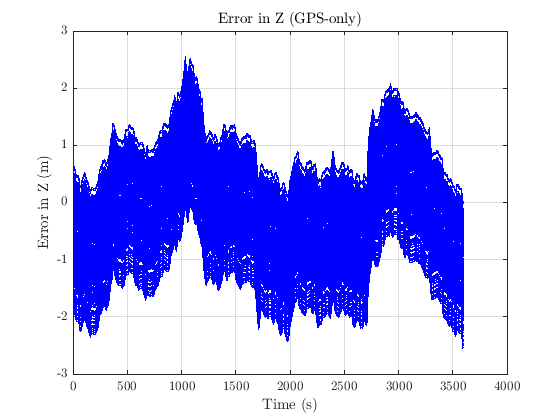} \\
    \end{tabular}
    \caption{Position errors in x, y, and z-coordinates using GNSS only.}
    \label{fig:gnss}
    \end{figure*}

The UKF is an advanced method for fusing GNSS and IMU data. Compared to the classic KF, it offers more accurate estimations of the state of a process that evolves non-linearly. The KF method relies on linear dynamics and measurements, whereas the UKF can handle non-linear systems without linearization.
The UKF's strength lies in its efficient use of the unscented transform. This involves selecting a minimal set of sample points around the mean, accurately capturing the state distribution's mean and covariance. These points are then propagated through the non-linear system, preserving the distribution's properties more effectively than linearization and enhancing the UKF's robustness.
In contrast, the standard KF may not be accurate enough when dealing with significant non-linearity, as it relies on linear approximations. In contrast, the UKF approach assumes that estimating a probability distribution is simpler than estimating a non-linear function.
Generally, the UKF implementation involves the prediction and measurement steps. 

{\bf Prediction Step:}
At time step {\em t}, the prediction step involves calculating the sigma points and updating the process by predicting mean and covariance \cite{noureldin2012fundamentals}. 
The sigma points can be calculated using:
\begin{equation}
\bar{X}_{t-1} = \begin{bmatrix} x_{t-1} & x_{t-1} \pm \sqrt{(n + \kappa)} P_{t-1} \end{bmatrix},
\end{equation}
where $\bar{x}_{t-1}$ denotes the sigma points of the state vector $\mathbf{x}$ at the prior time step $t - 1$. The state vector $\mathbf{x}$ has a dimension of $n$. The spread of the sigma points is defined by $\kappa = \alpha^{2}(n + \gamma) - n$. The parameter $\alpha$ determines the spread, while $\gamma$ is a secondary scaling factor, typically set to one. The initial condition must be known $x_{0} \sim N(x_{0}, P_{0})$.

  Then, the update process continues:
  \begin{equation}
\begin{aligned}
    \bar{x}_{t|t-1} &= f \left( x_{t-1}, u_{t-1} \right) \\
    x_{t}^{-} &= \sum_{i = 0}^{2n} W_{i}^{m} \bar{x}_{i, t|t-1} \\
    P_{t}^{-} &= \sum_{i = 0}^{2n} W_{i}^{c} \left( \bar{x}_{i, t|t-1} - x_{t}^{-} \right) \left( \bar{x}_{i, t|t-1} - x_{t}^{-} \right)^{\top} + Q,
\end{aligned}
\end{equation}

where $\bar{x}_t$ represents the predicted mean, $\bar{P}_t$ represents the predicted covariance, and $W_i^m$ and $W_i^c$ are the weights for the mean and covariance, respectively, associated with the i-th sigma point, as defined by \cite{wan2000unscented}:

\[W_0^m = \frac{\kappa}{n + \kappa}\]
\[W_0^c = \frac{\kappa}{n + \kappa} + (1 - \alpha^2 + \beta)\]
\[W_i^m = W_i^c = \frac{1}{2(n + \kappa)}, \quad i = 1, 2, \ldots, 2n\]

where $\beta$ is a parameter that allows for incorporating prior knowledge about the distribution of the state vector $\mathbf{x}$. In Gaussian distributions, the optimal value for $\beta$ is 2 \cite{wan2000unscented}.

{\bf Measurement Step:}
This step involves calculating the sigma points followed by the measurement update.
The sigma points calculation is:
\begin{equation}
    \bar{X}_{t} = 
    \begin{bmatrix}
        \bar{x}_{t} & \bar{x}_{t} \pm \sqrt{n + \kappa} P_{t}
    \end{bmatrix}
\end{equation}
where $\bar{x}_{t}$ represents the predicted mean from the time update at time $t$ and $\bar {P}_{t}$ denotes the predicted covariance. Then, the measurement update is done using the following equation.
\begin{equation}
\bar{Y}_{t} = g(\bar{X}_{t})
\end{equation}

\begin{equation}
\bar{y}_{t} = \sum_{i = 0}^{2n} W_{i}^{m} \bar{Y}_{i, t}
\end{equation}

\begin{equation}
P_{y_{t}} = \sum_{i = 0}^{2n} W_{i}^{c} (\bar{Y}_{i, t} - \bar{y}_{t}) \cdot (\bar{Y}_{i, t} - \bar{y}_{t})^{T} + R
\end{equation}

\begin{equation}
P_{x_{t}y_{t}} = \sum_{i = 0}^{2n} W_{i}^{c} (\bar{X}_{i, t} - \bar{x}_{t}) \cdot (\bar{Y}_{i, t} - \bar{y}_{t})^{T}
\end{equation}

\begin{equation}
K_{t} = P_{x_{t}y_{t}} \cdot \left( P_{y_{t}} \right)^{-1}
\end{equation}

\begin{equation}
v_{t} = y_{p, t} - \bar{y}_{t}
\end{equation}

\begin{equation}
x_{t} = \bar{x}_{t} + K_{t} \cdot v_{t}
\end{equation}

\begin{equation}
P_{t} = P_{t}^{-} - K_{t} \cdot P_{y_{t}} \cdot K_{t}^{T}
\end{equation}
where \(\bar{Y}_{t}\) signifies the sigma points projected through the measurement function \(h\),
while \(\bar{y}_{t}\) represents the predicted measurement based on weighted sigma points.
The predicted measurement covariance \(P_{y_{t}}\) and the state-measurement cross-covariance matrix \(P_{x_{t}y_{t}}\) 
are also derived from this process. \(K_{t}\) denotes the Kalman gain,
\(v_{t}\) is the innovation term, and \(x_{t}\) and \(P_{t}\) reflect the updated state and covariance at time \(t\).

Overall, the UKF is a more robust approach in situations where the Kalman Filter's linear assumptions are insufficient, especially in autonomous navigation and robotics, where accuracy in the face of non-linear behaviors is crucial.
\section{Experiments}
The experiment uses the KITTI GNSS and IMU dataset \cite{geiger2013vision}. 
The GNSS and IMU data are essential for tasks that involve vehicle pose estimation and tracking in autonomous driving. The GNSS provides geographic location information, while the IMU provides insights into the vehicle's movements through space, such as acceleration and orientation. Together, these data sets enable high-precision tracking and navigation, which is crucial in developing and testing autonomous systems. The dataset provides a dynamic and challenging real-world environment, with traffic scenarios recorded to reflect various driving conditions. It has served as a benchmark for many studies focusing on improving the accuracy and robustness of autonomous driving technologies. 

\begin{figure*}[!ht]
    \centering
    \begin{tabular}{ccc}
     \includegraphics[width=0.3\textwidth]{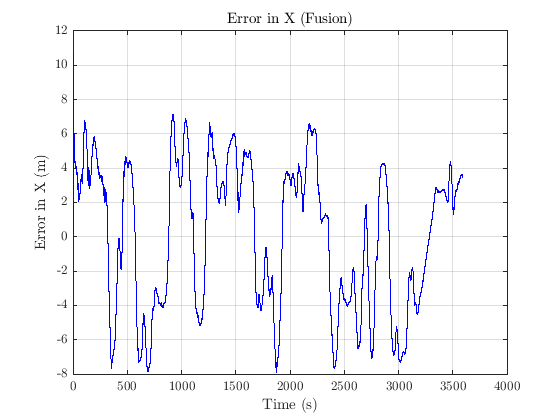} &
    \includegraphics[width=0.3\textwidth]{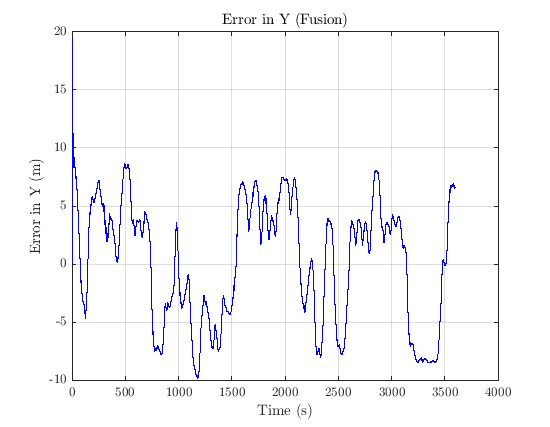} &
    \includegraphics[width=0.3\textwidth]{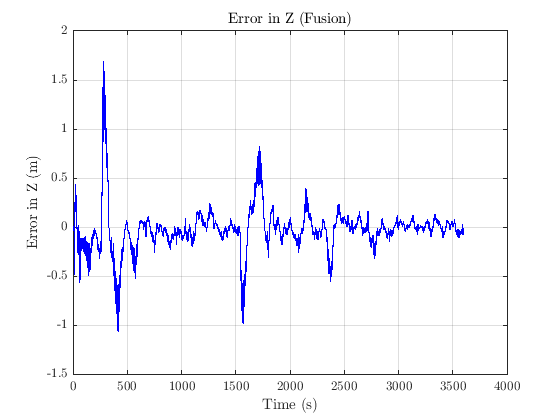} \\
    \end{tabular}
    \caption{Position errors in x, y, and z-coordinates using fusion of GNSS and IMU.}
    \label{fig:fusion}
    \end{figure*}

The experiment is conducted using a GNSS observation frequency of 1 Hz. Similarly, the IMU uses a 0.01 rad/s gyroscope, a 0.05 m/s\(^2\) accelerometer, a 0.000001 rad/s\(^2\) gyroscope bias, and a 0.0001 m/s\(^3\) accelerometer bias.
The position and velocity estimation using UKF results are demonstrated in the experiments. The UKF is a variant of the Kalman Filter that is more suitable for dealing with non-linear systems. It captures the mean and covariance of the probability distribution of the system state by passing carefully chosen sample points through the non-linearities.  The errors observed in the x, y, and z coordinates, as depicted in Figures \ref{fig:gnss} and \ref{fig:fusion}, serve as indicators of tracking accuracy over time. The GNSS-only results show the errors are larger than the fusion of GNSS and IMU data for navigation, as shown in Fig. \ref{fig:gnss}. The fusion figure, Fig. \ref{fig:fusion}, for the x and y directions reveals the GNSS frequency that provides the most precise location estimates, characterized by a lower mean error and minimal fluctuations around zero. On the other hand, the z-coordinate aims to identify the frequency that yields the smallest and most uniform error, highlighting a better tracking performance in three-dimensional space despite the inherent higher precision of GPS in the horizontal plane. Overall, these errors are essential in assessing the effectiveness of the fusion system.

\begin{figure*}[!ht]
    \centering
    \includegraphics[width=5in]{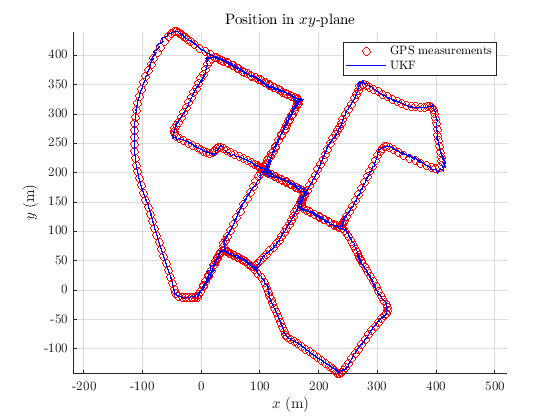}
    \caption{Positions in meters at GNSS frequency of 1 Hz.}
    \label{fig:pos}
\end{figure*}

Fig. \ref{fig:pos} demonstrating the position estimation in the xy-plane illustrates that the UKF estimated track closely aligns with the GPS measurements. The degree of tightness of the UKF track around the GPS measurements indicates the accuracy of the filter's performance.  Similarly, the RMSE across the X, Y, and Z axes are calculated to show the fusion model's robustness. 
The RMSE using GNSS-only shows  13.214, 13.284, and 13.363 for the X, Y, and Z axes, respectively. The fusion of the GNSS and IMU reduces the RMSE  to 4.271, 5.275, and 0.224 for the x, y, and z axes, respectively, as shown in Table \ref{tab:result}. For real-time implementation, certain challenges, such as sensor calibration and synchronization, need to be considered when incorporating sensor fusion technology into autonomous vehicles. Developing strong software algorithms, ensuring adequate hardware infrastructure, and conducting extensive testing in real-world scenarios are essential to addressing these challenges. This is crucial to guaranteeing autonomous vehicles' accuracy, reliability, and safety.

\begin{table}[!ht]
\centering
\caption{Comparison of RMSE for GNSS-only and GNSS-IMU fusion.}
\label{tab:result}
\begin{tabular}{llll}
\hline
\multirow{2}{*}{Method} & \multicolumn{3}{l}{RMSE}                                          \\ \cline{2-4} 
                        & \multicolumn{1}{l}{X-axis} & \multicolumn{1}{l}{Y-axis} & Z-axis \\ \hline
                        & \multicolumn{1}{l}{}       & \multicolumn{1}{l}{}       &        \\ 
    GNSS &13.214&13.284&13.363\\ 
    GNSS-IMU&4.271&5.275&0.224\\
    \hline
\end{tabular}%
\end{table}

\section{Conclusion}
The fusion model leveraging the UKF employs the KITTI dataset for validation, combining real-time IMU data with GNSS absolute positioning to refine autonomous vehicle navigation. The IMU's real-time motion data and the GNSS's absolute positioning capabilities are integrated to estimate precise and reliable poses. This is crucial to helping individuals remain aware of their surroundings and perform important navigation-related tasks, especially when faced with obstacles that might interfere with GNSS signals. This work highlights the UKF's superiority in processing nonlinear sensor outputs and reinforces the fusion model's potential for enhancing the reliability and efficiency of autonomous vehicle systems. Sensor fusion typically combines data from multiple sensors to improve navigation. It is also crucial during practical applications for maintaining position and orientation estimates if one of the sensors fails, allowing the navigation process to continue without interruption.
Adding LiDAR or radar sensors to autonomous vehicle navigation systems can improve the precision and dependability of position and orientation estimates by enhancing GNSS and IMU fusion. LiDAR's high-resolution 3D mapping capabilities enable accurate obstacle detection and localization, while radar's long-range detection and weather-resilience qualities enhance overall robustness. Fusing these sensors with GNSS and IMU can enhance the navigation system's reliability, ensuring continuous operation even during sensor failure. This leads to improved safety and navigation accuracy across various environments and conditions. Additionally, carefully developing existing filtering techniques, such as EKF, or customizing existing filters may improve performance. 

\end{document}